\documentclass[11pt]{article}

\usepackage[a4paper,margin=1in]{geometry}
\usepackage{microtype}
\usepackage[numbers,sort&compress]{natbib}
\usepackage{subcaption}
\usepackage{longtable}
\usepackage{graphicx}
\usepackage{tabularx}
\usepackage{multirow}
\usepackage{booktabs}
\usepackage{ragged2e}
\usepackage{fancyhdr}
\usepackage{doi}

\begin{document}

\title{Conceptualising an Initial Design Space for Guidance in Digital Physical Activity Support}

\author{
Faith Young$^{1,2}$ \and
Markus Tatzgern$^{3}$ \and
Alexander Meschtscherjakov$^{1}$ \and
Jan David Smeddinck$^{2}$
}

\date{}

\renewcommand{\thefootnote}{\fnsymbol{footnote}}
\footnotetext[1]{University of Salzburg, Salzburg, Austria}
\footnotetext[2]{Ludwig Boltzmann Institute for Digital Health and Prevention, Salzburg, Austria}
\footnotetext[3]{Salzburg University of Applied Sciences, Puch, Austria}
\renewcommand{\thefootnote}{\arabic{footnote}}

%
%

\maketitle

\begin{abstract}
Providing guidance is frequently referenced as a key capability of digital health interventions targeting physical activity, yet the term remains poorly defined and inconsistently applied. Existing work often conflates guidance with related constructs such as personalisation, feedback, or persuasion, limiting both theoretical clarity and design progress. This paper conceptualizes an initial design space of guidance in the context of digital physical activity support. We define guidance for physical activity as situated, action-oriented support that scaffolds users’ embodied engagement in physical activity. Drawing on literature from behaviour change, human–computer interaction, embodied cognition, and digital health, we outline a design space that characterises guidance along multiple dimensions: scope, purpose, timing, context, modality, embodiment, adaptivity, autonomy, and affective quality. By offering a structured vocabulary and conceptual foundation, this work aims to support more coherent research, comparisons, and responsible design of digital health interventions featuring guidance for physical activity support.
\end{abstract}

\section{Introduction}
Physical inactivity and sedentary behaviour are major global public health challenges, contributing substantially to the burden of non-communicable diseases such as cardiovascular disease, type 2 diabetes, cancer, and depression, as well as increased mortality worldwide \cite{ruegsegger2018, elkirat2024scoping}. Despite well-documented health benefits of regular physical activity (PA) and guidelines recommending at least 150 minutes of moderate-intensity PA per week, approximately one-third of adults fail to meet these targets \cite{who2020, elkirat2024scoping, Feil2023intention}. 

Beyond awareness of the benefits of PA, a central challenge lies in translating intentions into sustained behaviour. Although intention strongly predicts PA, nearly half of individuals with strong intentions do not act on them, creating the intention-behaviour gap \cite{sheeran2016intention, Feil2023intention, jorke2025gpt}. Recent ecological momentary assessment research shows that momentary intention, shaped by contextual and affective factors, is the proximal driver of PA enactment \cite{haag2025contextual}. This suggests that barriers to PA are often situational and dynamic; interventions that increase motivation alone may be insufficient if individuals lack support for acting in the moment.

Digital health interventions (DHIs) have shown promise in supporting PA at scale through mobile applications, wearable sensors, and web platforms \cite{kassim2025effect, jorke2025gpt}, primarily by enhancing motivation via goal-setting, self-monitoring, feedback, and persuasive prompts \cite{michie2011behavior}. While such approaches can increase engagement and short-term outcomes, many digital PA interventions continue to exhibit high attrition and limited long-term impact \cite{elkirat2024scoping, perski2017conceptualising}. Likely because DHIs mainly target the motivational component of behaviour change, while less attention has been paid to how systems can scaffold action in dynamic, real-world contexts \cite{smeddinck2025generativeai}.

The Motivation–Analysis–Guidance (MAGnify) framework ~\cite{smeddinck2016gamesforhealth, smeddinck2025generativeai, smeddinck2019}  identified guidance as a structurally underexplored area in health applications, observing that the field disproportionately focused on motivational design while treating guidance and analysis as secondary concerns. The present paper elaborates on this observation by proposing an initial structured design space for guidance that the original framework called for but did not provide. Despite progress in potential components and building blocks of guidance, such as Just-in-time adaptive interventions (JITAIs) exemplifying a shift toward context-sensitive, action-oriented support by delivering assistance at moments of high receptivity and situational relevance \cite{nahum2017jitai, riley2011health}, the notion of \textit{guidance} itself remains poorly defined in digital PA literature \cite{herold2022goingdigital, michie2011behavior}. Terms such as feedback, instruction, coaching, and nudging are often used interchangeably with guidance, creating conceptual ambiguity and hindering systematic research, evaluation, and design \cite{hekler2016advancing, oinas2009persuasive, perski2017conceptualising, klasnja2012healthcare}. Without a clear definition, it is difficult to assess whether an intervention is truly supporting users’ action, or merely providing information or motivational content.

This paper addresses this gap by proposing a definition and an initial design space characterisation of guidance for digital physical activity support. Building on insights from behaviour change theory \cite{michie2011behavior}, human–computer interaction, and digital health research, we structure guidance across multiple dimensions: scope, purpose, timing, context, modality, embodiment, adaptivity, autonomy, and affective quality. Our contribution is twofold: we (1) propose a working definition that clarifies guidance and distinguishes it from related concepts such as personalisation, feedback, and persuasion, and (2) introduce a structured design space that categorises how guidance can be understood, crafted and delivered, offering a shared conceptual language for researchers and designers. 

We position the design space as an initial conceptual contribution intended to support discussion, comparison, and future empirical refinement. By making guidance explicit as a design concern, this work lays a foundation for interventions that better support users in translating intentions into action, and facilitates systematic comparison, evaluation, and design of guided digital PA interventions.

\section{Conceptual Background}
\subsection{Digital Health Interventions for Physical Activity}
Digital interventions for PA have proliferated over the past two decades, leveraging mobile phones, wearable devices, and web platforms to deliver behaviour change support at scale. These systems commonly employ behaviour change techniques such as goal setting, self-monitoring, feedback, and reminders, often grounded in established theories including social cognitive theory, self-determination theory, and self-regulation models \cite{carver1982control, michie2011behavior, deci2000self}.

Systematic reviews indicate that digital PA interventions can produce small to moderate improvements in activity levels, particularly in the short term \cite{kassim2025effect}. However, effects often diminish over time, with high attrition rates and declining engagement posing persistent challenges \cite{eysenbach2005law, perski2017conceptualising}. These findings suggest that while digital technologies can increase access to PA support, sustaining behaviour change remains difficult.

\subsection{Guidance in Adjacent Domains}
The concept of guidance has been studied in a variety of related fields, each offering perspectives that inform the design of digital health interventions. In education and intelligent tutoring systems, guidance is commonly framed as scaffolding: providing learners with structured support to accomplish tasks they could not complete independently, with the goal of gradually fostering autonomy \cite{wood1976role}. In coaching and rehabilitation, guidance encompasses the provision of real-time instructions, corrective feedback, and motivational cues to support skill acquisition and adherence \cite{winstein2014stroke}. Similarly, in human–robot interaction, guidance refers to mechanisms by which robots assist humans in task execution, often blending autonomous actions with human decision-making to achieve shared goals \cite{fong2003survey}.

Across these domains, guidance is characterised by its \textit{action-oriented nature}, \textit{emphasis on timeliness} and \textit{situational relevance}, and its role in \textit{supporting performance and learning}. These principles provide a useful foundation for conceptualising guidance in the domain of PA. Also clear from these fields is when guidance is absent or insufficient, individuals often fail to translate intention into action, despite having the requisite skills or motivation. This gap is particularly pronounced in PA contexts \cite{haag2025contextual}. Recent ecological momentary assessment research demonstrates that momentary intention is the strongest predictor of PA enactment, yet its translation into behaviour is highly sensitive to contextual and affective fluctuations \cite{haag2025contextual}. Without timely, situated support and guidance, these momentary barriers can prevent intentions from being acted upon, contributing directly to the intention–behaviour gap.

The MAGnify framework positions guidance as the action-en\-abling layer between motivation and behaviour \cite{smeddinck2016gamesforhealth, smeddinck2025generativeai}. While motivational interventions target the ‘why’ of activity, guidance addresses the \textit{how, when, and where}, scaffolding action within real-world constraints. From this perspective, insufficient guidance may represent a missing explanatory layer in digital PA interventions: systems may increase motivation or awareness, yet still fail to support action when barriers are situational, bodily, or time-sensitive. Conceptualizing guidance as a distinct construct therefore provides both a theoretical and practical rationale for designing interventions that actively support behaviour enactment, rather than relying solely on motivation or informational content.

\subsection{Guidance in Digital Health and Physical Activity}
Within digital health, guidance is frequently invoked but rarely defined. Many interventions focus mainly on motivational aspects. Guidance is often mentioned in passing as “coaching” or “instruction,” yet it is typically not articulated as a distinct concept or operationalised in a systematic way \cite{klasnja2012healthcare, michie2011behavior}. This conceptual ambiguity limits comparability of interventions that claim to provide guidance, identification of which forms of guidance are most effective, and the design of systems that genuinely scaffold user action rather than merely informing or motivating it.

Moreover, as digital PA technologies increasingly incorporate AI and AR, the design space for guidance is expanding rapidly. Adaptive systems can provide in-the-moment, context-aware, and embodied support, raising both opportunities and ethical considerations such as autonomy, transparency, and dependency. Without a structured conceptualisation, it is difficult to evaluate or design these systems in a comprehensive and responsible manner.

\section{Defining Guidance in the Context of Physical Activity}
Building on insights from behaviour change theory, human–computer interaction, and related domains such as coaching and rehabilitation \cite{wood1976role}, we propose the following working definition of guidance:

\begin{center}
\textit{Guidance for digital physical activity is situated, action-oriented support that assists users in enacting physical activity through timely, context-sensitive, and embodied interventions.}
\end{center}

In this framing, a feature counts as guidance when it establishes an actionable relation between system support and user activity. A tailored workout plan, performance graph, or motivational prompt may contribute to guidance, but only becomes guidance when it helps the user decide what to do, how to do it, when to act, or how to adjust action in context. The above definition foregrounds four essential characteristics of guidance:

\textit{Action-Oriented:} Guidance focuses on enabling users to perform PA behaviours directly, bridging the gap between intention and action. Unlike personalisation, which primarily shapes the content of recommendations, guidance scaffolds doing, offering instructions, cues, or corrections in real time \cite{Feil2023intention}.

\textit{Situated:} Guidance is sensitive to the context in which behaviour occurs, including the physical environment, social setting, and users’ current capabilities. For example, an outdoor running app may adjust guidance based on weather, terrain, or proximity to safe routes \cite{russell2025digital}.

\textit{Embodied:} Guidance for PA involves support for the body as it engages in movement, whether through visual demonstrations, haptic feedback, or posture corrections. Embodied guidance acknowledges that PA is inherently experiential and sensorimotor, requiring more than abstract instructions \cite{dourish2001action}.

\textit{Timely and Adaptive:} Guidance is delivered at moments when it can influence action, and may adapt dynamically to user performance, progress, or situational changes. This distinguishes guidance from static instructional content or generic motivational messages \cite{gabarron2024human}. However, it is worth noting that guidance may still concern macro aspects of PA support (cf. Table 2 and section 4.2.1). In such cases \textit{timely and adaptive} still pertains to when and how such elements of guidance are delivered in the respective moments and scaffolding where they occur.

\subsection{Distinguishing Guidance from Related Concepts}
Guidance often combines elements of personalisation, feedback, instruction, and persuasion, but it is not reducible to any one of them. Table 1 summarises key distinctions between guidance and four commonly referenced related concepts: personalisation, feedback, instruction, and persuasion.

\begin{table*}[ht]
\centering
\begin{tabularx}{\textwidth}{|l|X|X|X|X|}
\hline
\textbf{Concept}   & \textbf{Primary Focus}                          & \textbf{Temporal Orientation}       & \textbf{Modality Examples}                             & \textbf{Relation to Action}                                           \\ \hline
Guidance           & Enabling action and scaffolding behaviour       & In-the-moment or contextually timed & AR overlays, voice cues, haptics, video demonstrations & Directly supports doing; tries to bridge intention–behaviour gap              \\ \hline
Personalisation \cite{smeddinck2016personalised}    & Tailoring content to user characteristics       & Often pre-activity or static        & Recommended workouts, difficulty levels, scheduling    & Shapes relevance of content; does not necessarily scaffold action     \\ \hline
Feedback \cite{hattie2007feedback}           & Informing user of performance or outcomes       & Post-action                         & Progress bars, metrics, graphs                         & Retrospective; may inform future action but not immediate execution   \\ \hline
Instruction \cite{schmidt2011motor}        & Providing knowledge about how to perform tasks  & Can be pre-activity or stepwise     & Text guides, videos, manuals                           & Supports understanding, not necessarily in-situ enactment             \\ \hline
Persuasion/Nudging \cite{oinas2009persuasive} & Motivating behaviour through psychological cues & Can be pre-, in-, or post-activity  & Notifications, badges, social prompts                  & Encourages behaviour, but does not provide detailed execution support \\ \hline
\end{tabularx}
\caption{Overview of guidance and related concepts in digital health interventions}
\end{table*}

While guidance may overlap with these constructs, its defining feature is direct, situated support for action, particularly during or immediately surrounding the performance of PA. Unlike feedback or personalisation, guidance is inherently operational: it helps users \textit{do} rather than merely \textit{know} or \textit{want}. Unlike persuasion or nudging, guidance prioritises skilful, safe, and embodied enactment rather than behavioural influence alone.

\section{Exploring the Design Space in Digital Physical Activity Support}
Building on the definition articulated above, we explore a multi-dimensional design space of guidance for digital PA interventions. Our design space identifies nine key dimensions along which guidance can be characterised: purpose, scope, timing, context, modality, embodiment, adaptivity, autonomy,and affective quality.

\subsection{Development of the Design Space}
The design space was developed through a purposive conceptual synthesis rather than a systematic review. We used the term guidance as an entry point and iteratively examined adjacent constructs in behaviour change, JITAIs, motor learning, coaching, embodied interaction, context-aware systems, persuasive technology, and adaptive user interfaces. We placed emphasis on the ACM DL as a search portal and used snowball search and discovery to find additional relevant works. Candidate dimensions were retained when they met three criteria: they described a recurring design choice, they shaped how support is delivered during or around PA, and they helped distinguish guidance from adjacent constructs such as feedback, instruction, personalization, or persuasion. The proposed dimensions therefore do not claim exhaustiveness but provide an initial vocabulary for analysing and designing guidance in digital PA support. While these dimensions are often examined independently, they are rarely integrated into a unified conceptual framework.



\begin{figure*}
    \centering
    \includegraphics[width=0.7\linewidth]{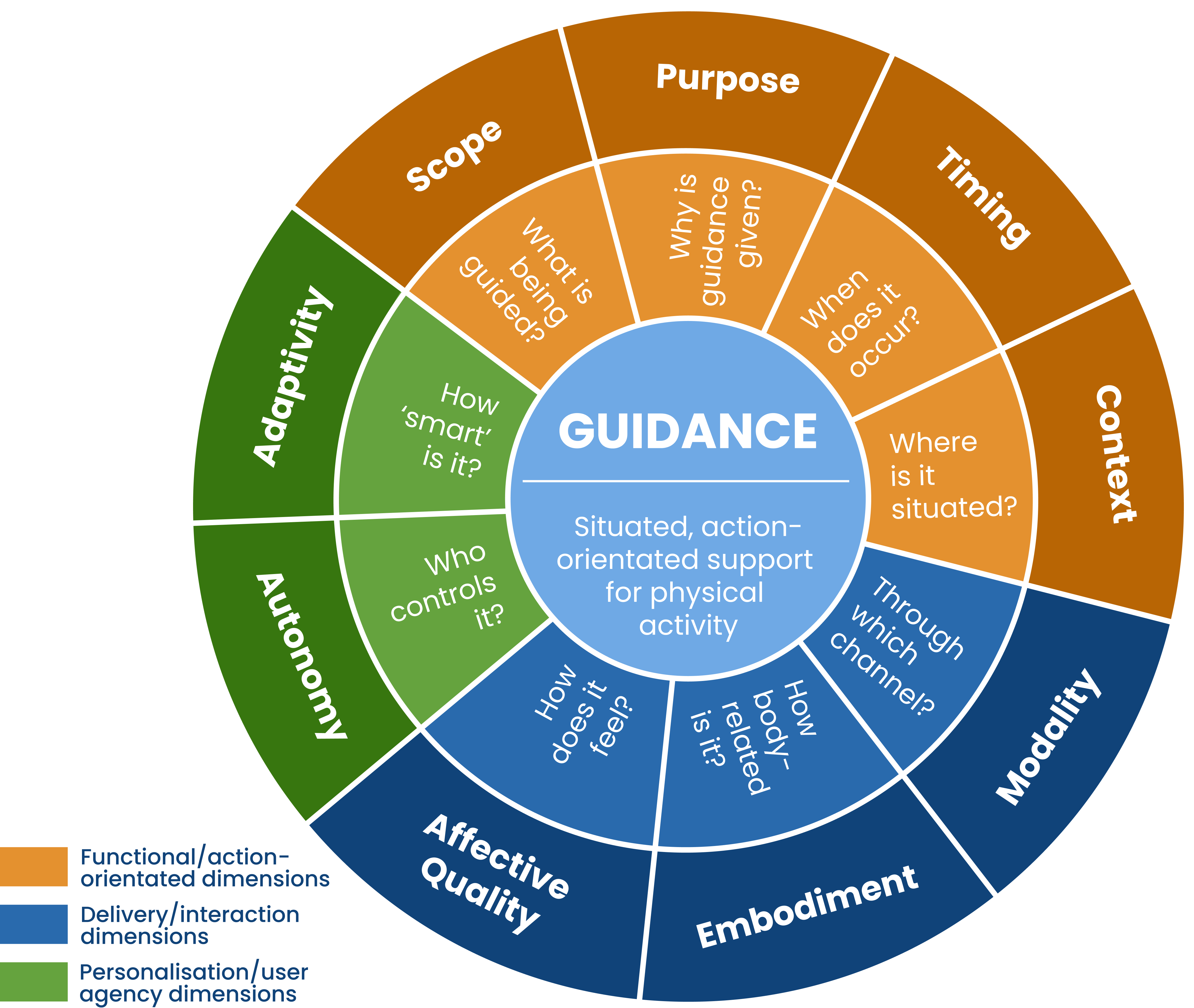}
    \caption{Guidance Design Space for Digital Physical Activity}
    \label{fig:design space visual}
\end{figure*}

The resulting design space characterisation represents a conceptual synthesis, organising established constructs into a coherent structure specifically focused on guidance in PA. This integrative framing enables systematic analysis, comparison, and design of guided PA interventions, while remaining grounded in established theoretical and empirical work. In this manner, we treated the design space as a sensitizing framework: its purpose is to make design choices visible and comparable, not to prescribe an optimal configuration of guidance. The design space can be seen in Figure \ref{fig:design space visual} and Table \ref{tab:design-space-questions} and dimensions are explained in further detail below.

\subsection{Design Space Dimensions}

\begin{table*}[ht]
\centering
\begin{tabularx}{\textwidth}{|l|l|X|}
\hline
\textbf{Design Question} & \textbf{Dimension} & \textbf{Range / Sub-aspects} \\ \hline
What is being guided? & Scope & Micro (motor execution) $\rightarrow$ Meso (session/activity management) $\rightarrow$ Macro (trajectory/pathway progression) \\ \hline
Why is guidance given? & Purpose & Scaffold action $\cdot$ Correct errors $\cdot$ Support reflection $\cdot$ Enable progression \\ \hline
When does it occur? & Timing & Pre-activity $\rightarrow$ In-the-moment $\rightarrow$ Post-activity; Continuous $\leftrightarrow$ Episodic \\ \hline
Where is it situated? & Context & Environment-aware $\cdot$ Socially embedded $\cdot$ Body-state-aware; Degree of situational responsiveness \\ \hline
Through which channel? & Modality & Visual $\cdot$ Auditory $\cdot$ Haptic $\cdot$ Multimodal; selection constrained by movement context and attentional demands \\ \hline
How body-related is it? & Embodiment & Abstract/symbolic $\rightarrow$ Body-referenced $\rightarrow$ Directly embodied \\ \hline
How 'smart' is it? & Adaptivity & Static $\rightarrow$ Rule-based $\rightarrow$ AI-driven; includes degree of context-sensitivity and personalisation \\ \hline
Who controls it? & Autonomy & System-driven $\leftrightarrow$ Negotiated $\leftrightarrow$ User-driven; includes transparency and overrideability \\ \hline
How does it feel? & Affective Quality & Encouraging $\leftrightarrow$ Neutral $\leftrightarrow$ Corrective (tone); Affectively blind $\rightarrow$ Affect-aware $\rightarrow$ Affect-adaptive (responsiveness); Persistent $\leftrightarrow$ Fading $\leftrightarrow$ Designed absence (presence dynamics) \\ \hline
\end{tabularx}
\caption{Guidance design space structured around nine dimensions and corresponding design questions. Arrows ($\rightarrow$) indicate continua; dots ($\cdot$) indicate non-ordered categories; double arrows ($\leftrightarrow$) indicate spectra with meaningful intermediate positions.}
\label{tab:design-space-questions}
\end{table*}

The design space of guidance in PA can be characterised along nine interrelated dimensions. Together, these dimensions describe how guidance is delivered, when it occurs, what it supports, and how it relates to user autonomy and embodied experience. The nine dimensions are \textit{Purpose}, \textit{Timing}, \textit{Context}, \textit{Modality}, \textit{Embodiment}, \textit{Adaptivity}, \textit{Autonomy}, \textit{Scope} and \textit{Affective Quality}. The dimensions are framed with a design question and the sub-aspects are classified as either continual, non-ordered, or spectra. 

\subsubsection{Scope}
Guidance in PA can operate at different levels of granularity. At the micro level, guidance targets motor execution within a single movement, such as posture correction during a squat. At the meso level, it addresses session management, including exercise sequencing or intensity regulation across a workout. At the macro level, guidance may concern longer-term trajectory progression, such as periodised training plans or rehabilitation pathway navigation. These levels are not mutually exclusive, and a single system may provide guidance across multiple scales simultaneously. The distinction is necessary since design requirements for micro-level guidance (low latency, embodied feedback) differ considerably from those for macro-level guidance (longitudinal awareness, goal coherence, integration with clinical planning) \cite{smeddinck2016gamesforhealth}.

\subsubsection{Purpose}
The purpose of guidance refers to the functional role it plays in supporting PA behaviour. Prior work across behaviour change and motor learning distinguishes between guidance aimed at initiating action, supporting execution, regulating performance, correcting errors, and facilitating reflection \cite{carver1982control, michie2011behavior, winstein2014stroke}.

For example, initiation-focused guidance may involve prompts or cues that support action planning and overcoming inertia. Execution and correction guidance align with coaching and motor learning principles, providing real-time feedback to regulate performance (e.g., pacing alerts during a run) or correct errors (e.g., haptic cues to adjust posture during a squat) \cite{schmidt2011motor}. Reflection-orientated guidance, such as post-activity summaries or feedback, supports learning and self-monitoring, which are critical for sustained behaviour change \cite{michie2011behavior}.

\subsubsection{Timing}
Guidance also varies in when it is delivered relative to PA. Temporal characteristics include pre-activity, in-the-moment, and post-activity guidance, as well as distinctions between continuous and episodic intervention. Continuous guidance provides ongoing support throughout the activity, such as a wearable device delivering real-time heart rate feedback. Episodic guidance, by contrast, occurs at discrete moments, for example, a smartphone notification reminding the user to perform a set of stretches before starting a workout.

This dimension is well established in the literature on JITAIs, which emphasise delivering support at moments of heightened receptivity or need \cite{nahum2017jitai}. In-the-moment guidance is particularly relevant for PA, where timely feedback can influence safety, performance, and engagement, whereas post-activity guidance often supports reflection and habit formation \cite{klasnja2019context, consolvo2008activity}.

\subsubsection{Context}
PA is inherently situated within specific environmental, social, and bodily contexts. Guidance systems therefore vary in the extent to which they are context-aware and contextually embedded.

The importance of context has long been recognised in both HCI and public health. Situated action theory highlights how behaviour emerges from interactions between individuals and their environments \cite{dourish2001action}, while ecological models of PA emphasise the role of environmental and social factors in shaping behaviour \cite{sallis2006ecological}. Context-aware systems aim to sense and respond to such factors, though prior work also cautions against over-automation that may undermine user agency \cite{barkhuus2003context}. 

\subsubsection{Modality}
The modality of guidance refers to the sensory channels through which guidance is delivered, including visual, auditory, haptic, and multimodal forms.

Multimodal interaction has been extensively studied in HCI, with evidence that different modalities afford different cognitive and attentional demands \cite{oviatt1999myths}. In PA contexts, visual guidance may be impractical during movement, making auditory or haptic modalities particularly valuable \cite{hoggan2008tactile}. Exertion interface research further highlights the need to align interaction modalities with bodily effort and movement constraints \cite{mueller2003exertion}.

\subsubsection{Embodiment}
Guidance can also be characterised by its degree of embodiment, ranging from abstract or symbolic representations to body-referenced feedback and directly embodied guidance.

This dimension draws on theories of embodied cognition and embodied interaction, which argue that cognition and action are deeply intertwined with bodily experience \cite{lakoff1999philosophy, dourish2001action}. In motor learning research, feedback that is closely coupled to bodily movement is often more effective than abstract instruction alone \cite{schmidt2011motor}. Treating embodiment as a continuum allows for more nuanced analysis of how guidance interfaces with the moving body.

\subsubsection{Adaptivity}
Guidance systems differ in the extent to which they adapt to users and contexts, ranging from static, pre-defined guidance to rule-based adaptation and AI-driven personalisation.

Adaptive user support has a long history in HCI \cite{oppermann1994adaptive}, while recent advances in mobile and wearable sensing have enabled increasingly sophisticated personalisation in health applications \cite{klasnja2012healthcare}. AI-driven guidance extends these approaches by enabling dynamic adaptation based on user behaviour, performance, and context. However, increasing system intelligence also raises concerns about transparency, predictability, user control, and ethical accountability \cite{amershi2019guidelines}.

\subsubsection{Autonomy}
Guidance also varies in how it distributes control and autonomy between the system and the user. This includes the extent to which guidance is optional, overrideable, and transparent.

Self-determination theory highlights autonomy as a key determinant of motivation and adherence in health behaviours \cite{deci2000self}. Within HCI, human-centred and value-sensitive design approaches emphasise the importance of user control, informed consent, and alignment with user values, particularly in AI-mediated systems \cite{friedman2006value, schneiderman2020human}. In the context of PA, overly controlling guidance risks undermining intrinsic motivation or fostering dependency.

\subsubsection{Affective Quality}
The tone and emotional register of guidance influence whether it is received as supportive, corrective, or controlling. Self-determination theory suggests that autonomy-supportive communication can support internalised motivation, while pressuring language may undermine it \cite{deci2000self, ng2012sdt}. In digital PA systems, affective quality therefore concerns not only whether guidance is encouraging, neutral, or corrective, but whether its tone, timing, and persistence fit the user's state and activity context. Recent work on LLM-generated health behaviour messages further suggests that contextual acknowledgement and perceived personal relevance can shape how users experience adaptive support over repeated interaction \cite{hofer2026personality}. We therefore characterise affective quality along three sub-aspects: tone, affective responsiveness, and presence dynamics (cf. \cite{wood1976role}).

\subsubsection{Integrating Dimensions}
These nine dimensions are interdependent. For example: Real-time AR guidance (modality + embodiment) may be highly adaptive (adaptivity) and aimed at execution and correction (purpose), requiring careful autonomy management. Post-activity reflection (time) may use abstract visualisation (modality/embodiment) and have low adaptivity, primarily supporting learning rather than immediate action.

Taken together, these dimensions provide a structured way to conceptualise guidance in PA as a multi-faceted phenomenon that spans behavioural and technological considerations. While each dimension is grounded in prior work, their integration into a unified design space enables more systematic design and evaluation of guided PA interventions.

\section{Discussion}
This paper set out to address a conceptual gap in digital physical activity support research by foregrounding guidance as a distinct and under-theorised design concern. While prior work has extensively explored personalisation, contextualisation, and behaviour change techniques; guidance has often been treated implicitly, embedded within features such as prompts, feedback or recommendations, rather than explicitly conceptualised. While prior frameworks have been instrumental in categorising intervention components, they offer limited vocabulary for describing how systems support embodied action as it unfolds \cite{michie2011behavior, torning2009persuasive}.

By proposing a design space characterisation that articulates multiple dimensions of guidance, this paper offers a clear definition of guidance in the context of digital PA as well as a potential means of systematically analysing and comparing interventions. While the framework has not yet been empirically applied, it offers a structured lens for examining differences not only in what interventions deliver, but in how they guide users across dimensions such as timing, modality, context, adaptivity, and autonomy. For example, two digital PA interventions might both encourage daily exercise, but one could provide real-time haptic cues during movement (high embodiment, in-the-moment guidance) while the other delivers daily text reminders (low embodiment, delayed guidance). Using the framework, researchers could map these interventions along the nine dimensions to identify patterns, gaps, or trade-offs in guidance design. This is particularly important in the context of PA, where the effectiveness, safety, and experience of an intervention may hinge more on timing, modality, and context rather than informational content alone \cite{schmidt2019motor, mueller2003exertion}.

\subsection{Reframing Guidance in Digital Physical Activity}
One key contribution of this work is the reframing of guidance from a purely informational or motivational mechanism to a relational and situated form of support. Unlike static content delivery, guidance actively shapes how users initiate, perform, regulate, and reflect on PA. This distinction is particularly important in the context of embodied action where timing, modality, and bodily engagement critically influence both experience and outcomes.

By articulating guidance across multiple dimensions, the proposed design space highlights that guidance is not a binary feature, but a design space with meaningful trade-offs. For example, in-the-moment, embodied guidance may enhance performance and safety, yet also risks increased cognitive load or perceived intrusiveness. Conversely, post-activity reflective guidance may support learning but lack immediate behavioural impact. Recognising these tensions allows designers and researchers to reason more explicitly about design choices rather than defaulting to generic solutions.

\begin{figure*}[t]
    \centering
    \includegraphics[width=\linewidth]{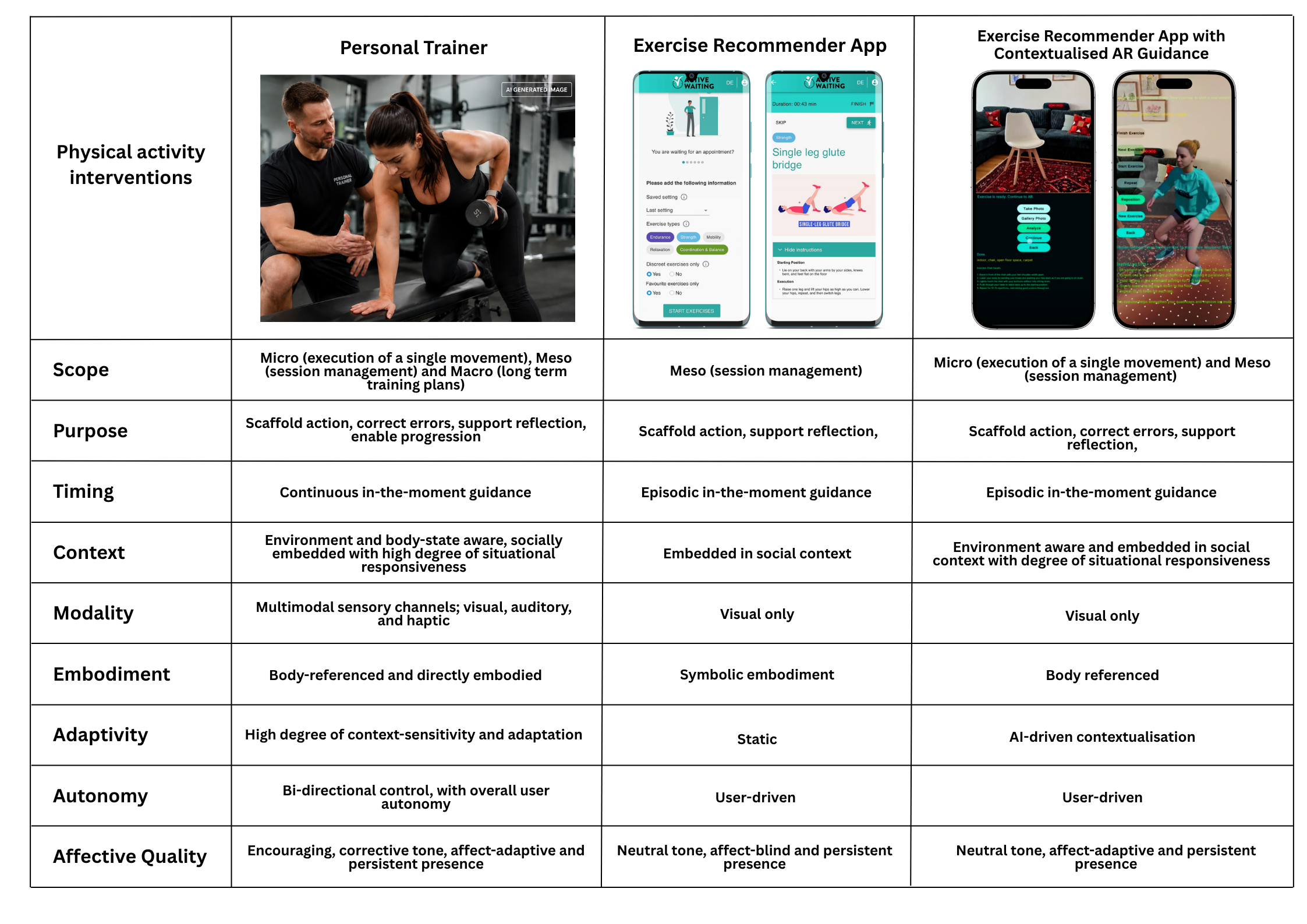}
    \caption{Illustrative mapping of three physical activity support configurations onto the proposed design space.}
    \label{fig:design space comparison}
\end{figure*}

\subsection{Clarifying Relationships to Existing Frameworks}
The proposed design space connects to the MAGnify framework introduced by Smeddinck \cite{smeddinck2016gamesforhealth, smeddinck2019}, which structures games for health around three interdependent conceptual lenses: Motivation, Analysis, and Guidance and has since been employed with digital health interventions more generally. Accordingly, our work can be understood as an elaboration of the guidance lens, extended beyond game-based interventions into broader digital PA support. Importantly, MAGnify highlights that guidance and motivation are not independent: how guidance is delivered shapes autonomy, competence, and intrinsic motivation. By generalising the guidance lens from game-specific mechanics to non-game PA interventions, the present work extends the prior conceptualisation into contexts with less inherent structure and greater reliance on external sensing and AI-driven adaptation.

Accordingly, the proposed design space complements rather than replaces existing behaviour change frameworks. For example, while the Behaviour Change Technique Taxonomy identifies what techniques are used, it does not specify how these techniques are operationalised in action \cite{michie2011behavior}. Behaviour change models tend to emphasise motivation, intention, and self-regulation, while HCI research focuses on interaction modalities and system adaptivity. Motor learning and coaching literature, in contrast, foreground execution quality and embodied feedback. Similarly, Just-in-Time Adaptive interventions emphasise temporal optimisation but typically treat intervention components as messages or prompts rather than embodied guidance processes \cite{nahum2017jitai}. The guidance dimensions articulated here make these operational differences explicit. This added resolution may help explain variability in intervention outcomes that cannot be accounted for by technique selection alone, aligning with calls for more fine-grained analyses of intervention delivery and user experience \cite{perski2017conceptualising}.

By combining these perspectives, the design space can provide a shared conceptual vocabulary that can support interdisciplinary research and design. In particular, it highlights how guidance in PA must account simultaneously for behavioural intent, interaction design, and bodily action. This integrative framing may help bridge gaps between research communities that address similar problems using different conceptual tools.

\subsection{Embodiment and Situated Action as Key Elements of Guidance}
A key contribution of this work is foregrounding guidance as an embodied and situated phenomenon. PA unfolds through movement, perception, and effort, making the body a central site of interaction rather than a peripheral input mechanism. Insights from embodied cognition and interaction suggest that guidance delivered through movement is processed differently than abstract information, often relying on sensorimotor coupling rather than reflective reasoning \cite{dourish2001action, lakoff1999philosophy}. The proposed characterisation of the design space has embodiment and modality dimensions which highlight how guidance can range from symbolic representations to body-referenced cues. This perspective aligns with motor learning research demonstrating that timely, task-relevant feedback can scaffold performance while excessive or poorly timed guidance may impair skill acquisition or autonomy \cite{schmidt2019motor}. By explicitly accounting for these factors it supports more nuanced design decisions in activity-focused systems.
 



\subsection{Implications for Design and Evaluation of DHIs}
The design space encourages designers to treat guidance as a design element, prompting explicit decisions about purpose, timing, modality, and autonomy. It also provides a structured way to compare systems that may appear similar on the surface but differ substantially in how they guide users. For researchers, the design space offers a foundation for more nuanced evaluation. Rather than measuring only behavioural outcomes such as adherence or step count, studies can examine how different configurations of guidance affect user experience, perceived autonomy, learning, and long-term engagement. The dimensions can also complement reporting frameworks such as TIDieR and CONSORT-EHEALTH by specifying how support is delivered as guidance during or around embodied activity, not only what intervention components are present \cite{hoffmann2014better, eysenbach2011consort}. 
To illustrate how the design space can be used, Figure \ref{fig:design space comparison} maps three indicative configurations: human personal training, a conventional exercise recommender, and a contextualised AI-supported recommender. The mapping is illustrative rather than evaluative; its purpose is to show how guidance configurations can be described and compared.

By conceptualising guidance as a multi-dimensional and embodied phenomenon, this work outlines a foundational framework for understanding how digital systems support PA. The proposed design space does not seek to prescribe optimal solutions, but to enable more deliberative, reflective, and responsible design of guided PA interventions. In doing so, it opens new avenues for research at the intersection of digital health, HCI, and behaviour change.

\subsection{Limitations}
This work is primarily conceptual and therefore has several limitations. First, the proposed design space has not yet been empirically validated. Future work should examine how the dimensions interact in practice and which configurations are most appropriate across different physical activity contexts and user groups. Second, the framework is derived primarily from Western literature and may not fully capture culturally specific understandings of guidance, authority, or physical activity. Finally, while intended to be comprehensive, the design space is neither exhaustive nor static. As digital health technologies continue to evolve through AI, extended reality, and emerging sensing technologies, additional dimensions or refinements may become necessary.




\section{Conclusion}
This paper proposed a multi-dimensional design space for guidance in digital PA interventions, addressing a critical conceptual gap. Drawing on behaviour change, HCI, motor learning, and embodied cognition, the framework provides a structured lens for understanding how, when, and to what extent guidance supports action.

The design space articulates nine interrelated dimensions: scope, purpose, timing, context, modality, embodiment, adaptivity, autonomy, and affective quality. These dimensions provide a structured vocabulary for describing, comparing, and designing guidance beyond existing accounts centred primarily on motivation or behaviour change techniques.

Beyond conceptual clarity, the design space can inform the design of more effective and context-sensitive interventions, support systematic evaluation of guidance configurations, and foster interdisciplinary research. While empirical validation remains a future step, this work lays the foundation for understanding guidance not merely as a feature or add-on, but as a central mechanism for supporting behaviour change and embodied action in everyday life.


\bibliographystyle{plainnat}
\bibliography{sample-base}


\end{document}